\documentstyle[12pt,aaspp4,epsfig]{article}
\begin{document}
\title{Structure of Sagittarius A* at 86 GHz using VLBI Closure Quantities}

\author{
S. Doeleman\footnote{MIT Haystack Observatory, Off Route 40 Westford, MA 01886},
Z.-Q. Shen$^{2,3}$\addtocounter{footnote}{1}\footnotetext{National Astronomical Observatory, Mitaka, Tokyo 181-8588 JAPAN}
\addtocounter{footnote}{1}\footnotetext{ASIAA, PO Box 1-87, Nankang, Taipei, TAIWAN},
A.E.E. Rogers$^1$,
G.C. Bower\footnote{NRAO, PO Box 0, 1003 Lopezville Rd., Socorro, NM 87801},
M.C.H. Wright\footnote{University of California, Astronomy Dept. 601 Campbell Hall, Berkeley, CA 94720},
J.H. Zhao\footnote{CFA, 60 Garden St., Cambridge, MA 02138},
D.C. Backer$^5$,
J.W. Crowley$^1$,
R.W. Freund\footnote{NRAO, 949 North Cherry Ave, Tucson AZ 85721},
P.T.P. Ho$^6$,
K.Y. Lo$^3$,
D.P. Woody\footnote{California Institute of Technology, Owens Valley Radio Observatory, Big Pine, CA 91125}
}
\begin{abstract}
At radio wavelengths, images of the compact radio source Sagittarius A* (Sgr A*)
in the Galactic Center are scatter broadened with a $\lambda^2$ dependence due
to an intervening ionized medium.  We present VLBI observations of Sgr A* at 86
GHz using a six station array including the VLBA antennas at Pie Town, Fort
Davis and Los Alamos, the 12~m antenna at Kitt Peak and the millimeter arrays at
Hat Creek and Owens Valley.  To avoid systematic errors due to imperfect antenna
calibration, the data were modeled using interferometric closure information.
The data are best modeled by a circular Gaussian brightness distribution of FWHM
$0.18 \pm 0.02$ mas.  The data are also shown to be consistent with an
elliptical model corresponding to the scattering of a point source.  The source
structure in the N-S direction, which is less well determined than in the E-W
direction due to the limited N-S $(u,v)$ coverage of the array, is constrained
to be less than 0.27 mas by these measurements.  These results are consistent
with extrapolations of intrinsic structure estimates obtained with VLBI at 7~mm
wavelength assuming the intrinsic size of Sgr A* has a greater dependence than
$\lambda^{0.9}$ with wavelength.
\end{abstract} 



\newpage

\section{Introduction}

Sgr A* is the compact radio source in the center of the Galaxy and is believed
to be powered, as are extra-Galactic AGN, via accretion onto a massive black
hole.  Evidence for this scenario solidified recently with the stellar proper
motion studies of Eckart \& Genzel (1997) and Ghez et al. (1998) which show
velocity dispersions of $\sim1400\,\mbox{km}\;s^{-1}$ near the Galactic center
implying a $2.6\times10^6M_\odot$ central mass within a $10^{-6}\;\mbox{pc}^3$
volume.  Upper limits on the proper motion of Sgr A* itself (Backer \& Sramek
1999, Reid et al. 1999), constrain its mass to be greater than $\sim
2\times10^4M_\odot$.  These results would seem to rule out stellar type
phenomena as models for the radiative output from Sgr A* which spans radio to
x-ray wavelengths.  Estimates of the luminosity of Sgr A* (Zylka et al. 1992)
though, are only a small fraction ($\sim 10^{-5}$) of the Eddington limit for a
$2.6\times10^6M_\odot$ black hole (Quataert et al. 1999).  Physical models that
reproduce the observed Sgr A* emission spectrum in the context of mass estimates
above include scaled AGN models (Falcke \& Biermann 1999), advection dominated
spherical accretion flows (ADAF) (Narayan, Yi \& Mahadevan 1995) and
cyclo-synchrotron emission from spherical accretion (Melia 1992).

Direct measurement of the intrinsic size of Sgr A* is of great importance in the
discrimination of different emission mechanisms thought to operate in this
source.  In the radio bands, such measurements are complicated by scattering
through the inhomogeneous ionized ISM which broadens images.  Past VLBI
observations ranging in wavelength from 6~cm to 1.3~cm, observe an elliptical
source with axial ratio 0.53 oriented E-W whose dimensions vary with $\lambda^2$
as expected for scattering of an unresolved source by electron density
irregularities (Lo et al.  1985, Jauncy et al. 1989, Marcaide et al. 1992,
Alberdi et al. 1993).  Similar elliptical scattering disks are observed for OH
masers located within 25 arc minutes of Sgr A* (Frail et al. 1994).  Such
non-circular scattering implies an anisotropy in the underlying density
inhomogeneities that may arise due to ordered magnetic fields in the region.

Using 7~mm wavelength VLBI, Lo et al. (1998) reported a departure from this
axial ratio of the scattering ellipse, a result consistent with independent
measurements by Bower \& Backer (1998).  The apparent departure from the
scattering, if confirmed, would be strong evidence for intrinsic structure in
Sgr A*.  A possible elongation of Sgr A* at 7~mm was first reported by Krichbaum
et al. (1993) who modeled Sgr A* with two VLBI components separated by 3 mas along
a PA of $-25^\circ$.  Later 7~mm VLBI work by Backer et al. (1993) and Bower and
Backer (1998) found no evidence for such structure.  Lo et al. (1998) interpret
their 7~mm result as intrinsic structure consisting of an elongation of $0.44
\pm 0.09$ mas with position angle of $-10^\circ$.  The reported statistical
significance of this result is 5 standard deviations, but there are additional
systematic errors in the calibration (Bower et al. 1999).  VLBI observations of
Sgr A* at 3.5~mm wavelength, which are much less affected by interstellar
scattering, can provide important verification of the 7~mm results.  Intrinsic
structure of Sgr A*, if visible at 7~mm, should be readily observable at shorter
wavelengths as long as its size does not decrease faster than $\lambda^{2}$.

Previous size estimates for Sgr A* from VLBI at 3.5~mm wavelength have been
obtained using arrays of two (Rogers et al. 1994) and three (Krichbaum et al.
1998) antennas.  The sparse data sets of Rogers et al. (1994) and Krichbaum et
al. (1998) were modeled with circular Gaussian brightness distributions and
yielded sizes of $0.15\pm 0.05$ mas and $0.19\pm0.03$ mas respectively,
consistent with a $\lambda^2$ extrapolation of the scattering size at this
wavelength.  There remain some differences in the literature as to the exact
power law which best characterizes the relation between scattering size and
wavelength (Backer et al. 1993, Rogers et al. 1994, Krichbaum et al. 1998).
Differences in the number and weighting of past size measurements are largely
responsible for the variation in fitted relations.  In this paper, we will adopt
the relations cited by Lo et al. (1998) which yield a size of $1.43\;\lambda^2$
mas for the major axis and $0.76\;\lambda^2$ mas for the minor axis where
$\lambda$ is measured in cm.  These result from nearly simultaneous observations
reduced in a uniform manner.

The limited baseline coverage and sensitive dependence on a-priori antenna
calibration of previous arrays have not yet allowed a definitive estimate of
intrinsic structure at shorter wavelengths, though a detection of Sgr A* at 1.3
mm wavelength has been reported (Krichbaum et al. 1998).  This paper reports on
new VLBI observations made at 3.5~mm wavelength with an array of six antennas that
allowed use of closure quantities to model the Sgr A* structure.  We explore two
different methods of model fitting using the closure amplitude information.
Closure amplitudes and phases provide structural information without the need to
calibrate the VLBI array (Readhead et al. 1980).

\section{Observations}

\renewcommand{\thefootnote}{\fnsymbol{footnote}}
\setcounter{footnote}{1}

Observations of Sgr A* ($\alpha(\mbox{J2000})=17^h45^m40^s.045,\;
\delta(\mbox{J2000})=-29^\circ00'27.9''$; Rogers et al. 1994) at 86 GHz were
made by the Coordinated Millimeter VLBI Array\footnote{Support for the
Coordinated Millimeter VLBI Array work at the Haystack Observatory is provided
under a grant from the NSF to the Northeast Radio Observatory Corporation}
(CMVA) in April 1999.  Fringes were obtained to the KittPeak 12m antenna (K),
the Owens Valley mm-Array (O), the BIMA array at Hat Creek (H), and the
NRAO\footnote{The National Radio Astronomy Observatory is a facility of the
National Science Foundation operated under cooperative agreement by Associated
Universities, Inc.} VLBA sites at Los Alamos (L), Pie Town (P), and Fort Davis
(F), where the symbols to be used in this paper for each site are in
parentheses.  Sgr A* and continuum calibrators NRAO 530, OV-236 were observed on
both 16 and 18 April 1999, but technical problems at the VLBA sites reduced the
array to just 3 antennas on the first day.  The data presented here were
observed during the second day in which Sgr A* was detected on baselines to all
sites.  A total bandwidth of 56 MHz consisting of 14 separate 4 MHz channels was
observed for 360 second duration scans.  SiO masers towards VX Sgr were observed
at both 43~GHz and 86~GHz for pointing and calibration respectively.  The data
were processed on the Mark IIIA correlator at Haystack Observatory with an
integration time of 1 second.

\section{A-Priori Calibration}

Accurate a-priori amplitude calibration of VLBI visibilities relies completely
on the successful determination of telescope sensitivity as a function of time
and elevation using previously established gain information.  At the low
elevations required for Sgr A* observations, mm-wave dishes such as the Kitt Peak
12~m can be accurately characterized, but gains for the VLBA antennas, which are
not yet optimized for 86~GHz observations, are difficult to determine.  Phasing at
the mm arrays also adds a degree of gain variation that depends sensitively on
time variable atmospheric conditions.  Two independent a-priori calibration
techniques were applied to the array.  

The SEFD (System Equivalent Flux Density) was first determined for each antenna
as a function of time using system temperature (Tsys) measurements and
previously established gain-elevation curves.  VLBA antennas use internal noise
sources to determine Tsys, and these values were corrected for atmospheric
opacity effects.  The mm arrays and Kitt Peak use a vane absorber method to
measure Tsys and no opacity correction is required.  Explicit assumptions in
this type of array calibration are perfect pointing and stable gain curves at
all sites.  A faulty calibration source prevented Tsys from being recorded at
the VLBA Fort Davis site.  

A second calibration method made use of total power spectra obtained by auto
correlating VLBI data on the 86 GHz SiO maser source VX Sgr
($\alpha(\mbox{J2000})=18^h08^m04^s.04,\;
\delta(\mbox{J2000})=-22^\circ13'26.90''$; Wright et al. 1990) to generate 112
channel spectra in a 4MHz bandpass.  A template to the maser line structure from
the Kitt Peak auto correlations was calibrated assuming a flat gain curve and a
sensitivity of 0.025~K/Jy.  This template spectrum was then fitted to total
power spectra extracted from all VX Sgr observations at all array sites to
determine SEFD of each antenna as a function of time.  This method of
calibration assumes perfect pointing only for the template spectrum, pointing
errors on all other VX Sgr observations will be corrected relative to the
template scan.  At BIMA, a phase calibration tone leaking into the bandpass
prevented calibration using the spectral template technique.

Fig.~\ref{fig:sefd}  shows the results of these two calibrations for all sites.
At four sites (Kitt Peak, OVRO, Pie Town and Los Alamos) both the spectral
template (filled circles) and gain curve (solid line) calibrations were used.
The Kitt Peak and OVRO sites show generally good agreement between the two
methods although it appears that the gain curve at Kitt Peak may have more
elevation dependence than originally assumed.  The gain curve derived SEFD at
Kitt Peak rises more slowly at lower elevations than the VX Sgr derived values.
For the two VLBA sites, correspondence between the two calibrations is
distinctly inferior.  In the elevation range $20^\circ$ to $30^\circ$, the gain
curve calibration is consistently below the VX Sgr SEFD values by as much as a
factor of 4 in the case of Los Alamos, and by a factor of 2.5 for Pie Town.  The
difference between the calibration schemes is probably due to pointing errors at
the VLBA sites which are given, along with their effects on VLBI amplitudes, in
Table ~\ref{tab:pointing}.  The loss factors given are 1\,$\sigma$ values, so
the observed scatter in the spectral template results is entirely consistent
with pointing errors as well.  At 86~GHz, the VLBA pointing is affected by known
systematic pointing offsets at the 20 arc second level due to ripple in the
azimuth track (Dhawan, private communication).  Both a-priori calibration
methods thus do not compensate for pointing errors on Sgr A* observations at the
VLBA antennas.  In order to avoid these difficulties in calibration, which are
likely not tractable at 3.5~mm, we chose to use only closure amplitudes in our
analysis which are not susceptible to calibration errors and are determined on
observations of Sgr A* itself.  

\section{Data Analysis}

The Sgr A* data was analyzed using the following procedure:

1]  Fringes were obtained at all sites on the calibration sources to determine
the calibration phases which allow coherent bandwidth synthesis of the baseband
channels.  In addition the calibration sources were used to estimate the
clock delays and clock rates for each station.

2]  A fringe search was conducted on Sgr A* to refine the delays and rates using
the Haystack Observatory Postprocessing System (HOPS) software.  This step used
interpolations of delay and rate from adjacent calibrator scans as well as
strong Sgr A* detections on sensitive baselines to constrain searches on weaker
baselines.  This method works even when there are very weak fringes on some
baselines, as long as there are sufficiently strong fringes on enough baselines
to reduce the search space (e.g. Alef \& Porcas 1986).  The data were also
imported into the NRAO AIPS software package where independent fringe searches
resulted in consistent delay and rate results.

3]  The noise bias corrected incoherent average of 10 second data segments was 
computed for each baseline using the refined clock delays from the 
constrained fringe search.  These data segments are short enough to avoid significant 
signal loss due to atmospheric path fluctuations.  The incoherent average
$\langle a\rangle$  (Rogers, Doeleman, Moran 1995) is 
\begin{equation}
\langle a\rangle=\left[\frac{1}{M}\sum_{s=0}^{M-1}\left(|a_s|^2-2{\sigma_c}^2\right)\right]^{\frac{1}{2}}
\end{equation}
where $a_s$ is the complex correlation from the coherent integration of each segment and $\sigma_c$  
is the noise bias
\begin{equation}
\sigma_c=\frac{1}{L}\left(2BT\right)^{-\frac{1}{2}}
\end{equation}
\begin{tabbing}
wherexxxxx\= \kill  
where\>B = 56 MHz \\
\>T = coherent integration = 10 sec \\
\>L = loss factor $\simeq$ 0.55 for digital MK3 1-bit correlation including fringe 
rotator loss \\
\>M = number of segments 
\end{tabbing}

4]  Closure phases, the sum of interferometric phases around a closed loop of
three baselines (Jennison 1958, Rogers et al. 1974), were estimated for each 360
second VLBI recording scan.  The complex valued bispectrum or triple product
(Rogers, Doeleman \& Moran 1995), whose argument is the closure phase, was
calculated for each 10 second data segment and averaged to a full scan length.
This avoided any coherence losses due to phase fluctuations over the length of
an entire 360 second VLBI observation.  Deviations of the closure phase from
zero on any triangle would indicate an asymmetric component in the structure of
Sgr A*, zero closure phases are the signature of a brightness distribution
symmetric about the origin.  For a brightness distribution dominated by a
central compact feature, closure phases are not expected to change significantly
on a 360 second time scale.  Closure phase plots in Bower \& Backer (1998) show
that even for a 2 component model of Sgr A* at 7~mm, closure phases vary by
$<5^\circ$ in a six minute period.  In the low signal to noise (SNR) regime, the
bispectral SNR is proportional to the product of the SNRs of each contributing
baseline and it can be shown (Rogers, Doeleman \& Moran 1995) that this SNR is
greater than that of the closure phase alone.  We employ a bispectral SNR cutoff
which ensures an SNR in each 10 second segment on each baseline of approximately
1.6.  This corresponds to a $15^\circ$ standard deviation in the closure phase.  

The measured closure phases are consistent with zero values.  Several
representative closure phase triangles whose high SNR enabled us to track their
values over multiple scans are listed here with their measured standard
deviations : FKP = $24^\circ\pm12$, FKL = $10^\circ\pm23$, KOL =
$11^\circ\pm22$, FKH = $3^\circ\pm20$, FKO = $-6.3^\circ\pm27$.  Subsequent
modeling was restricted to symmetric brightness distributions.

5]  A circular or elliptical Gaussian is fit to the amplitudes allowing separate
station- based gains for each 360 second scan.  This is similar to the commonly
used self-calibration technique for imaging interferometric data except that
self-calibration cannot search all possible brightness distributions.  Instead,
self-calibration converges iteratively on a consistent set of complex antenna
gains and an image by using a deconvolution method such as CLEAN.  The resulting
image is not unique (Schwarz 1978) due to the iterative nature of the imaging
process.  Alternatively, the method described below performs an exhaustive
search over all models in a restricted parameter space and guarantees that the
result is the most likely. 

The best fit model is found by minimizing $\chi^2$.  For a weighted least
squares,
\begin{equation}
\chi^2 = \sum_{ij}w_{ij}(t)\left\{a_{ij}(t)-m_{ij}e^{g_i(t)}e^{g_j(t)}\right\}^2
\label{eq:chisq}
\end{equation}
 
\begin{tabbing}
wherexxxxx\= \kill
where\>$m_{ij}(t)$is the model at time $t$ for baseline $ij$ \\
\>$a_{ij}(t)$ is the observed incoherent average of data segments \\
\>$e^{g_i(t)}$ is the gain for station $i$ for each scan at time t \\
and\>$w_{ij}(t) = 1/{\sigma_{ij}}^2(t)$ \\ 
where\>${\sigma_{ij}}^2(t)$ is the estimated variance in $a_{ij}(t)$
\end{tabbing}

The standard deviation in each noise bias corrected amplitude estimate in correlation 
units is 
\begin{equation}
\sigma=\sigma_cM^{-\frac{1}{2}}\left(\left(s^2+1\right)/s^2\right)^{\frac{1}{2}}
\end{equation}
	 
\noindent where $s$ is the coherent SNR for each segment.  Monte Carlo
simulations show this approximation to hold for a coherent SNR =$a/\sigma_c \geq
0.5$.  For example, a correlation of $3\times10^{-5}$ has a SNR = 0.5; with 10s
segments and M = 36, $\sigma\simeq 1.5\times10^{-5}$.  Baselines with
correlations below $3 \times 10^{-5}$ were not used in the model fit.

The summation is carried over all available baselines and scans.  For each scan
and each model the station 
gains are estimated by the weighted least squares (Bevington \& Robinson 1992,
see also Scharf 1991 for matrix notation)
solution of 
\begin{equation}
y=Ax+n
\end{equation}
given by
\begin{equation}
\hat{x}=\left(A^TwA\right)^{-1}A^Twy
\end{equation}

\begin{tabbing}
wherexxxxx\= \kill
where\>$x = [g_0, ... ,g_{L-1}]^T$ = gain vector for L sites which participate in scan \\
\>$w$ = diagonal weight matrix \\
\>$y$ = data vector \\
\>$A$ = steering (or curvature) matrix \\
\>$n$ = noise or error vector \\
\>$T$ = the transpose operator \\
\>$\hat{x}$ = weighted least squares estimate of $x$ \\
\>$^{-1}$ = the inverse operator 
\end{tabbing}

The $n$th element of the data vector is $y_n = \ln(a_n/m_n)$
where $n$ is the baseline number.  The design matrix elements are
\begin{tabbing}
xxxxx\= \kill
\>$A_{nk}$ \= = $1$ when $k = i$ or $k = j$ \\
\>\>= 0 otherwise
\end{tabbing}

where $k$ is the station number.  The use of logarithms linearizes the
estimation of the gains.  The search for a minimum $\chi^2$ using
Eq.~\ref{eq:chisq} requires separate gain solutions for each model tested since
the gain solutions depend on the model.  A 1-D Monte Carlo search is made for
the circular Gaussian with minimum $\chi^2$, while an elliptical Gaussian
requires a 3-D search through parameter space.  Pearson (1995) discusses the
least squares fitting of Gaussian models to VLBI data and the estimation of
errors.  Under the assumption of a Gaussian distribution of errors, the weighted
least squares approach, which simultaneously solves for the model and the
station gains, is equivalent to the maximum likelihood solution.  Using this
method we obtained the best fit circular and elliptical Gaussian models shown in
Table ~\ref{tab:mod} (models A and B).  

The resulting least squared station gains are directly compared with the
a-priori results in Fig.~\ref{fig:sefd}, represented by crosses.  In most cases,
the modeled station SEFD values are less than spectral template values at
comparable elevations.  This may be due to the proportionality of the modeled
SEFDs on the total flux density of Sgr A* which we have assumed to be 1.4~Jy
based on BIMA observations taken prior to the VLBI observations (Wright, private
communication).  If the flux of Sgr A* during our observing epoch was higher,
the derived station gains from the modeling would increase.

Figure~\ref{fig:model} compares the gain corrected visibility amplitudes with a
model for the best fit circular Gaussian.  The same data is recast as a plot of
correlated flux density as a function of baseline length in
Figure~\ref{fig:uv_amp}.  The fit to an elliptical Gaussian is not well
determined in the N-S direction owing to the reduced $(u,v)$ coverage in this
direction (see Figure~\ref{fig:uv}).  Errors in the circular and elliptical
Gaussian models were estimated from the size of the region in parameter space
corresponding to $\Delta\chi^2\leq1$.  In Figure ~\ref{fig:chisq} we have
plotted $\chi^2$ as a function of the FWHM of the circular Gaussian model.  We
explored a range of coherent averaging times from 4 to 60 seconds and found very
little change in our results from 4 to 20 seconds.  At 60 seconds ${\chi_\nu}^2$
was over 1.5 and the estimated error in the size of the best fit circular
Gaussian more than doubled.

We also modeled the data using closure amplitude quadrilaterals which are
insensitive to station gains.  The closure amplitude is defined as :
\begin{equation}
C_{\mbox{ijkl}} =
\frac{a_{\mbox{ij}}\;a_{\mbox{kl}}}{a_{\mbox{ik}}\;a_{\mbox{jl}}}
\end{equation}
a ratio of complex amplitudes around a closed quadrilateral loop of baselines.
The number of independent closure amplitudes is equal to the number of baselines
between all sites minus the number of sites so that for 6 sites there are 9
independent closure amplitudes.  Cornwell (1995) discusses the equivalence of
self-calibration and modeling of closure quantities in determining source
structure.  However, using $\chi^2$ from fitting the closure quadrilaterals to
estimate the errors in source parameters is not straightforward as the
quadrilaterals are not all independent.  Furthermore, in the low SNR case, the
formal error for each closure amplitude is difficult to determine (see e.g.
Trotter, Moran, Rodr\'{i}guez 1998, eqns. 4 and 5).


Based upon VLBA measurements between 6~cm and 7~mm wavelength, Lo et al. (1998)
have suggested that the 7~mm results show deviation of the minor axis size from
the $\lambda^2$ dependence implying an intrinsic size of $0.44\pm0.1$ mas along
PA=$-10^\circ$.  Since the scattering angles are smaller at 3.5~mm wavelength,
it is important to compare the constraints set by the 3.5~mm data to reasonable
extrapolations from the Lo et al. (1998) inferred intrinsic structure.  Model C
constitutes pure $\lambda^2$ interstellar scattering of a point source.  Model D
is the extrapolated apparent source size based on an intrinsic source size that
varies as $\lambda^{0.9}$ (Lo et al. 1999).  Model E represents the case in
which the 7~mm derived intrinsic structure does not vary with wavelength.
Estimates of apparent source size at 3.5~mm and 1.3~mm (Rogers et al. 1994,
Krichbaum et al. 1998) are consistent with some sort of wavelength dependent
intrinsic structure from 7~mm to 1.3~mm, but these results depend sensitively on
a-priori calibration.  The exact behavior of the Sgr A* intrinsic structure as a
function of frequency is thus largely uncertain with estimates for a power law
dependence ($\lambda^\alpha$) ranging from $\alpha=0.7$ to 1.9.  Table
~\ref{tab:mod} contains characteristics and fit results for models C, D and E.
For all elliptical models, the reported PA reflects the major axis orientation.

\section{Discussion}

The 3.5~mm observations are consistent within 2\,$\sigma$ of the elliptical
scattering model of a point source.  They are also consistent at the 3\,$\sigma$
level with a model combining the elliptical scattering disk with an intrinsic
structure extrapolated from the Lo et al. (1998) 7~mm estimates assuming a
$\lambda^{0.9}$ dependence.  Increasing the spectral index of this power law,
and consequently decreasing the intrinsic size, improves agreement with the
3.5~mm data as the observed size approaches the scattering size.  To further
illustrate the level of agreement of the 3.5~mm data with all models, we show in
Figure ~\ref{fig:clamp} a closure amplitude for the Fort Davis, Kitt Peak, OVRO,
Hat Creek quadrangle ($C_{\mbox{FKOH}}$) along with the same closure amplitude
values expected for models A through E.  Model E, shown as the solid line in the
lower panel of Figure ~\ref{fig:clamp}, provides the poorest fit to the data.
The dotted lines show the effects on model E if the N-S intrinsic structure from
Lo et al. (1998) is reduced or increased by their cited 1\,$\sigma$ errors.
Other quadrilaterals also show similar inconsistencies with model E.

It has been customary in VLBI studies of Sgr A* to model the brightness
distribution as an elliptical Gaussian given the scattering morphology seen at
longer wavelengths.  At shorter wavelengths ($\lambda \leq$ 7~mm), however, it is
not clear that the data warrant a model more complex than a circular Gaussian.
Model A in Table ~\ref{tab:mod} has the lowest $\chi^2$ of any model tested, but
it also contains two extra model parameters relative to Model B.  The
$F_\chi$-test (Bevington \& Robinson 1992) is the proper method to determine the
statistical significance of a drop in $\chi^2$ when additional model parameters
are added.  The result of such an analysis is based on trying to fit both
circular and elliptical Gaussian models to pure noise and calculating the
difference in $\chi^2$.  Given the number of data points in our observations, we
find that a $\Delta \chi^2=5$ between the two models occurs at least 12\% of the
time when fitting pure noise.  The better fit of model A, therefore, is only
significant at the $1.5\sigma$ level, and we conclude that the data presented
here are adequately fit by a circular Gaussian.  A similar test has not yet been
performed for previous 7~mm wavelength data sets (Lo et al. 1998, Bower \& Backer
1998), and such an analysis would be useful in determining the significance of
their results.

Looking for departures from extrapolated scattering sizes due to intrinsic
structure is difficult with current arrays.  Based on contours of $\Delta\chi^2
= 14$ we can limit the elongation in a position angle of $-10^\circ$ to less
than 0.27 mas.  At 7~mm the effects of intrinsic structure are small and imaging
can be influenced by calibration of the VLBA antennas.  Previously reported
measurements at 3.5~mm by Rogers et al. (1994) and at 1.4 mm by and Krichbaum et al.
(1998) have insufficient baselines to be able to rely on closure and are
consequently limited by calibration errors.  Closure phases derived by summing
interferometric phase over station triplets are generally consistent with zero
value.  This constrains possible brightness distributions to be symmetric about
the origin, a characteristic shared by all the models tested here.  Adding
secondary components to the Sgr A* structure produce marginally better fits to
the closure phase data, but not enough to justify the additional degrees of
freedom.  The closure amplitude data presented here show no evidence of
asymmetry.

Assuming the scattering and intrinsic structure add in quadrature, limits on the
intrinsic size for Sgr A* in position angles of $-10^\circ$ and $80^\circ$
are 0.25 mas and 0.13 mas respectively.  For a 1.4 Jy source, these limits
correspond to a minimum brightness temperature of $7\times10^{9}$K, nearly half
the value derived in Rogers et al. (1994).  The essential difference between the
two results is the assumed structure, which Rogers et al. (1994) models as a
circular Gaussian.  The limits derived here assume an elliptical model allowing
for an extension in the N-S direction and consequent increase in size.

\"{O}zel et al. (2000) have recently computed new ADAF models incorporating
generalized electron energy distributions that reproduce the observed Sgr A*
emission spectrum.  Their size estimate at 7~mm closely matches the size of the
N-S extension seen by Lo et al. (1998), but is much larger than the upper limit
to the intrinsic size set in the E-W direction.  The spherical ADAF models may
have limited applicability to Sgr A* if an intrinsic ellipticity or asymmetry is
verified at shorter wavelengths.  At 3.5~mm, the ADAF model (\"{O}zel et al.
2000) predicts an intrinsic size of 0.29 mas, well over the 3\,$\sigma$ upper
limit in the N-S direction derived here (Figure~\ref{fig:sgsize}).  In the AGN
model of Falcke \& Biermann (1999), the estimate of intrinsic size for Sgr A* at
7~mm is used to constrain parameters of the relativistic jet, and the emission
region size varies as $\lambda^{0.89\chi_\circ}$ where $\chi_\circ$ is a factor
related to the jet orientation.  An extrapolation to 3.5~mm gives a size of 0.24
mas, larger than the 3.5~mm E-W size limit but consistent with N-S limits.  

\section{Summary}

We have used amplitude closure information from a 86 GHz VLBI array to determine
a size estimate of Sgr A* without the need to calibrate the array with a-priori
information.  With an assumption of elliptical Gaussian structure, the data are
consistent with interstellar scattering of an unresolved source.  Our results
also do not exclude extrapolations from estimates of intrinsic size at 7~mm
wavelength as long as the intrinsic size evolves faster than $\lambda^{0.9}$.  A
best fit elliptical model does exhibit an elongation in the N-S direction, but
the improvement in fit over a circular Gaussian model was found to be of
marginal statistical significance.  The N-S $(u,v)$ coverage of the VLBI array
is sufficient to constrain an intrinsic extension in the $-10^\circ$ position
angle to be less than 0.25 mas.  Emission models that reproduce the spectrum of
Sgr A* predict sizes for Sgr A* that are at or above this size limit.

\section{Acknowledgements}

We thank the staff at participating observatories for help with the observations
and the staff of Haystack Observatory for correlating the data.  We also thank
Ramesh Narayan and Feryal \"{O}zel for making data on their ADAF models available to
us prior to publication.

\section{References}

Alberdi, A., Lara, L., Marcaide, J.M., et al. 1993, A\&A, 277, 1

Alef, W. \& Porcas, R.W. 1986, A\&A, 168, 365

Backer, D., \& Sramek, R.A. 1999, ApJ, 524, 805

Backer, D.C., Zensus, J.A., Kellerman, K.I., Reid, M., Moran, J.M. \& Lo, K.Y.
1993, Science, 262, 1414

Bevington, P.R. \& Robinson, D.K. 1992, Data Reduction and Error Analysis for the
Physical Sciences, 2nd edition (New York: Mc-Graw-Hill)

Bower, G.C., \& Backer, D.C. 1998, ApJ, 496, L97

Bower, G.C., Falcke H., Backer, D.C., \& Wright, M.C.H. 1999, in ASP Conference
Series 186, The Central Parsecs of the Galaxy, eds. H. Falcke, A. Cotera, W.J.
Duschl, F. Melia, \& M.J. Rieke, p. 80

Cornwell, T. 1995, in ASP Conf. Series 82, Very Long Baseline Interferometry
and the VLBA, eds. J.A. Zensus, P.J. Diamond, \& P.J. Napier, p. 39

Doeleman, S., Rogers, A.E.E., Backer, D.C., Wright, M.C.H., \& Bower, G.C. 1999,
in ASP Conf. Series 186, The Central Parsecs of the Galaxy, eds. H. Falcke, A.
Cotera, W.J. Duschl, F. Melia, \& M.J. Rieke, p. 98

Eckart, A., \& Genzel, R. 1997, MNRAS, 284, 576

Falcke, H. \& Biermann, P.L. 1999, A\&A, 342, 49

Ghez, A. M., Klein, B. L., Morris, M., \& Becklin, E. E. 1998, ApJ, 509, 678 

Jauncy, D.L., Tzioumis, A.K., Preston, R.A., et al. 1989, AJ, 98, 44

Jennison, R.C. 1958, MNRAS, 118, 276

Krichbaum, T.P., Zensus, J.A., Witzel, A. et al. 1993, A\&A 274, L37

Krichbaum, T.P., Graham, D.A., Witzel, A. et al. 1998, A\&A 278, L1

Lo, K.Y., Backer, D. C., Ekers, R. D., Kellermann, K. I., Reid, M. \& Moran, J.
M. 1985, Nature, 315, 124

Lo, K.Y., Shen, Z.-Q., Zhao, J.-H. \& Ho, P.T.P. 1998, ApJ, 508, L61

Lo, K.Y., Shen, Z.-Q., Zhao, J.-H. \& Ho, P.T.P. 1999, in ASP Conf. Series 186,
in ASP Conf. Series 186, The Central Parsecs of the Galaxy, eds. H. Falcke, A.
Cotera, W.J. Duschl, F. Melia, \& M.J. Rieke, p. 72

Marcaide, J.M., Alberdi, A., Bartel, N., et al. 1992, A\&A, 258, 295

Melia, F. 1992, ApJ, 387, L25
 
Narayan, R., Yi, I. \& Mahadevan, R. 1995, Nature, 374, 623

\"{O}zel, F., Psaltis, D. \& Narayan, R. 2000, ApJ, 541, 234

Pearson, T.J. 1995, ASP Conference Series v.82, pp 267-286

Quataert, E., Narayan, R. \& Reid, M.J. 1999, ApJ, 517, L101

Readhead, A.C.S., Walker, R.C., Pearson, Cohen, M.H. 1980, Nature, 285, 137

Reid, M. J., Readhead, A. C. S., Vermeulen, R. C., \& Treuhaft, R. N. 1999, ApJ,
524, 816

Rogers, A.E.E. et al., 1974, ApJ, 193, 293

Rogers, A.E.E. et al., 1994, ApJ, 434, L59

Rogers, A.E.E., Doeleman, S.S., Moran, J.M. 1995, AJ, 109, 1391

Scharf, L.L. 1991, Statistical Signal Processing : Detection, Estimation, and Time
Series Analysis (New York: Addison-Wesley)

Schwarz, U.J. 1978, A\&A, 65, 345

Trotter, A.S., Moran, J.M. \& Rodr\'{i}guez, L.F. 1998, ApJ, 493, 666

Wright, M.C.H., Carlstrom, J.E., Plambeck, R.L., \& Welch, W.J. 1990, AJ, 99, 1299

Zylka, P., Mezger, P.G. \& Lesch, H. 1992, A\&A, 261, 119
 
\begin{table}[p]
\begin{center}
\caption{Signal Loss Due to Pointing Errors at VLBA Sites \label{tab:pointing}}
\vspace{0.5cm}
\begin{tabular}{ccccc} \tableline\tableline
Antenna & rms pointing (arcsec) & loss factor 1 & loss factor 2 & loss factor 3\\
(1) & (2) & (3) & (4) & (5) \\
\tableline
VLBA-PT & 17 & 0.33  & 0.76 & 0.58 \\
VLBA-LA & 14 & 0.47 & 0.83 & 0.69  \\
VLBA-FD & 13 & 0.64 & 0.90 & 0.80 \\ \tableline
\end{tabular}
\end{center}

\tablecomments{Columns: (1) Site; (2) rms of the difference between consecutive
pointings performed every 20 minutes during the observations; (3) loss of
sensitivity due to the pointing errors: $\exp(-\ln(2)\; \mbox{error}^2/
\mbox{HWHM}^2)$ where HWHM is the half width at half maximum of the antenna
beam; (4) loss assuming the telescopes are properly pointed at the beginning of
each scan but drift to the pointing error values during a 20 minute scan; (5)
loss given by a mixture of the two cases.}

\end{table}

\begin{table}[p]
\begin{center}
\caption{Description and Fit of Models for Sgr A* \label{tab:mod}}
\begin{tabular}{lcccccccc} \tableline\tableline
Model & $\theta_{\mbox{maj}}$(mas) & $\theta_{\mbox{min}}$(mas) & PA (deg) &
$\chi^2$ & ${\chi_\nu}^2$
& $\Delta \chi^2$ & $\sigma$ & Comments  \\ 
(1) & (2) & (3) & (4) & (5) & (6) & (7) & (8) & (9) \\
\tableline
A & $0.34\pm 0.14$ & $0.17\pm 0.02$ & $22\pm 20$ & 97 & 1.19 & 0 & -
& Best Fit Elliptical \\
B & $0.18\pm 0.02$ & --- & --- & 102 & 1.25 & 5 & 1.5 & Best Fit Circular \\
C & 0.175 & 0.092 & 80 & 104 & 1.28 & 7 & 2 & Scattering Alone \\
D & 0.263 & 0.175 & -10 & 111 & 1.36 & 14 & 3 & Lo et al. model scaled \\
E & 0.44 & 0.175 & -10 & 247 & 3.03 & 150 & 10 & Lo et al. model unscaled \\
\tableline
\end{tabular}
\end{center} 

\tablecomments{Columns: (1) model name; (2) FWHM major axis; (3) FWHM minor
axis; (4) position angle of major axis (for elliptical models); (5) $\chi^2$ of
the model fit; (6) reduced $\chi^2$ obtained by dividing $\chi^2$ by the number
of baselines minus the rank of the matrix $A^TwA$ (see text); (7) $\chi^2$
difference from model A; (8) equivalent formal standard error with equivalent
confidence at the $\Delta\chi^2$ for the 3 degrees of freedom in an elliptical
Gaussian model; (9) model description.}

\end{table}

\newpage

\figcaption{Comparison of three calibration methods for the VLBI array.  Filled
circles mark System Equivalent Flux Density (SEFD) measured using auto spectra
on the line calibrator VX Sgr.  This method reflects the effects of pointing
errors.  The second calibration marked by crosses is the result of solving for
station gains in the model fitting process described in the text.  Similar
degrees of scatter in the gains found by the two methods suggest they are both
tracking similar variations.  Pointing errors tabulated in Table
~\ref{tab:pointing} probably account for most of the observed gain variation.
The third calibration used published gain curves and measured system
temperatures to calculate the SEFD and is shown as solid lines. \label{fig:sefd}
} 

\figcaption{$(u,v)$ coverage corresponding to all observations of Sgr A*.
\label{fig:uv}}

\figcaption{Calibrated visibility amplitudes for
each baseline between Fort Davis (F), Kitt Peak (K), Owens Valley (O), Hat Creek
(H), Pie Town (P) and Los Alamos (L).  Calibration of each antenna
was determined from a weighted
least squares search through possible circular Gaussian models of brightness
distribution.  The dotted lines show amplitude curves for the best fit circular 
Gaussian model (FWHM=0.18 mas).  Error bars are 3\,$\sigma$ and the flux density
scale for all baselines is the same.
\label{fig:model}} 

\figcaption{Calibrated visibility amplitudes as in Figure~\ref{fig:model} plotted
as a function of baseline length.  The solid line shows the best fit circular
Gaussian model and dotted lines show circular Gaussians that are $\pm1\,\sigma$
in size.  Error bars are identical to those shown in Figure~\ref{fig:model} and
are left off here for clarity.  The scatter in the data is reflective of the
errors.
\label{fig:uv_amp}}

\figcaption{$\chi^2$ vs the FWHM for a circular Gaussian model.
\label{fig:chisq}}

\figcaption{Measured closure amplitudes for the Fort Davis, Kitt Peak, OVRO, Hat
Creek quadrilateral along with closure amplitude curves for models A through E.
The upper panel compares the closure amplitudes for models A through D.  The
lower panel shows model E and two variants: model E is the solid grey curve, the
two dotted lines reflect 1\,$\sigma$ errors in the N-S intrinsic structure
at 43 GHz suggested by Lo et al. (1998).  \label{fig:clamp}}

\figcaption{Size measurements of Sgr A* from VLBI at multiple wavelengths.
Green squares are minor axis sizes and red dots are major axis sizes of best fit
elliptical models at wavelengths less than 3.5~mm from Lo et al. (1998).  Green
and red solid lines are $\lambda^2$ scattering relations given by Lo et al.
(1998).  The 3.5~mm upper limit to the size in the N-S direction from this work
is shown as a horizontal green line. The N-S observed size predicted from the
estimated 7~mm intrinsic structure (model D in Table 2) is the green filled
circle, almost exactly at the upper size limit.  Black triangles show sizes of
circular Gaussian models fit to the 3.5~mm data described in this work, and to
1.4~mm VLBI data (Krichbaum et al.  1998).  Hybrid ADAF model sizes from \"{O}zel,
Psaltis, and Narayan (2000) are added in quadrature to both the minor and major
axis scattering sizes and shown as the green and red dotted lines respectively.
The observed sizes predicted by the ADAF model exceed the 3.5~mm upper limits.
\label{fig:sgsize}}

\begin{figure}[p]
 \begin{center}
\epsfig{figure=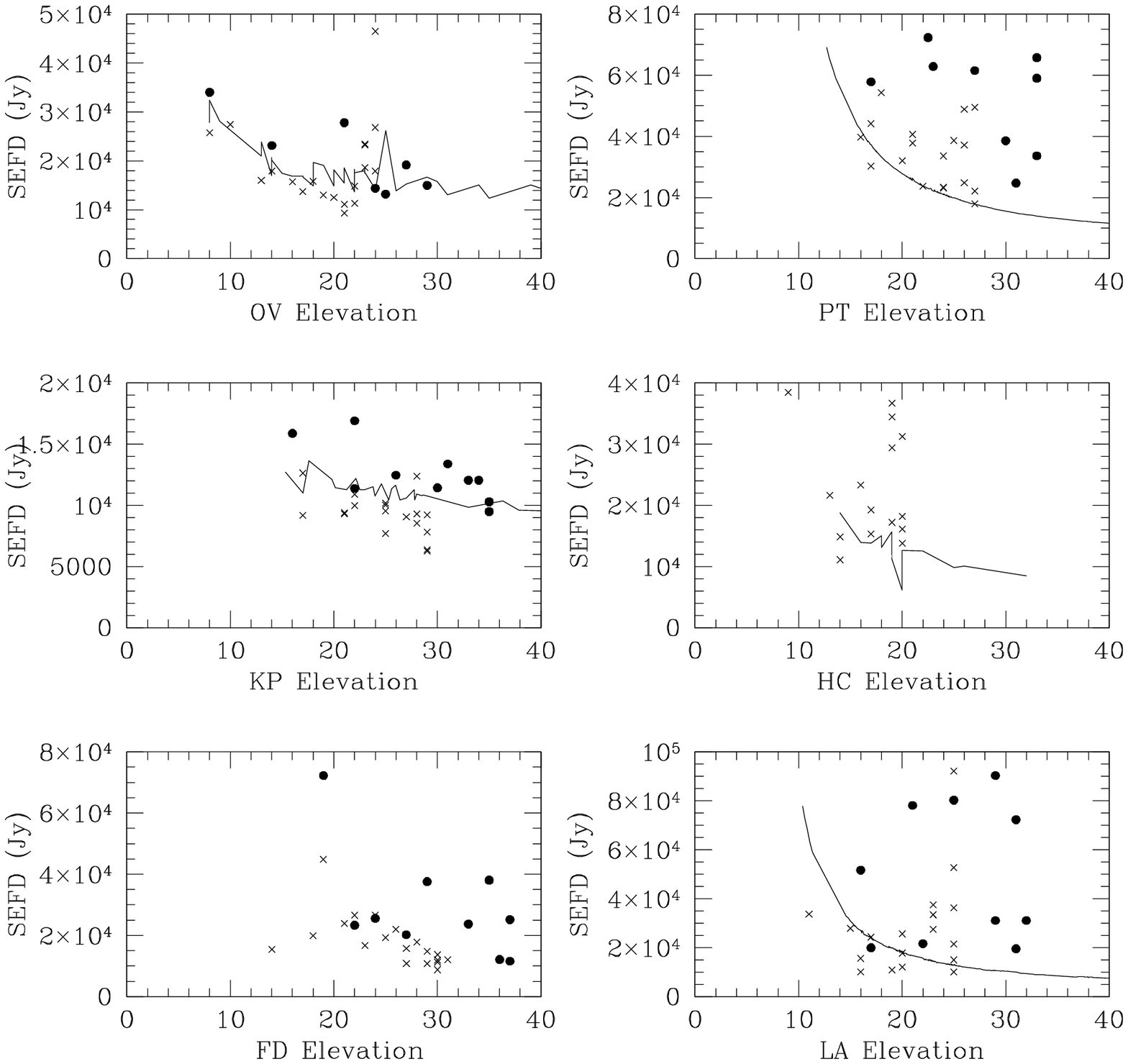,width=6.5in}
 \end{center}
\end{figure}

\begin{figure}[p]
 \begin{center}
\epsfig{figure=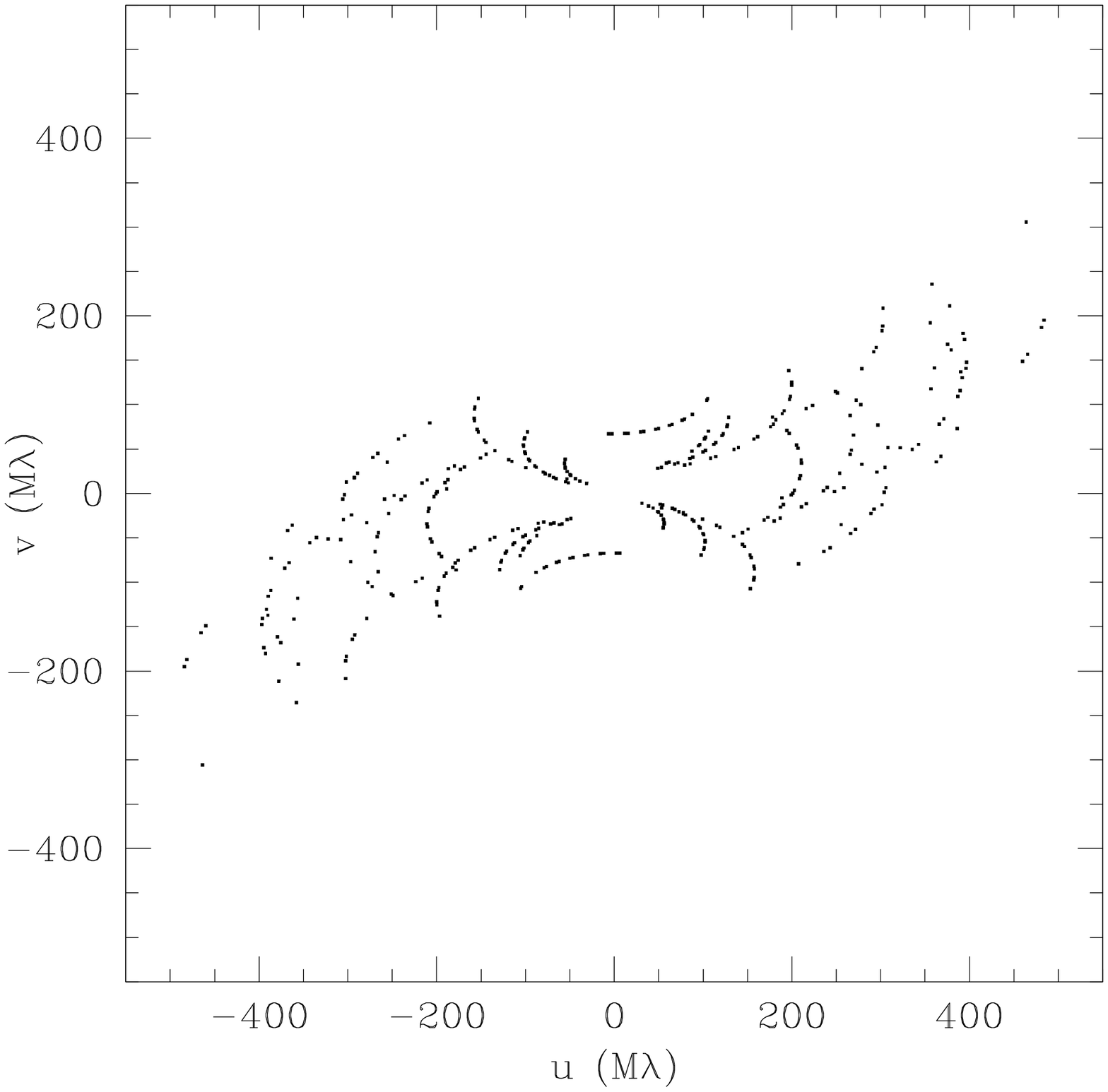,width=6.5in}
 \end{center}
\end{figure}

\begin{figure}[p]
 \begin{center}
\epsfig{figure=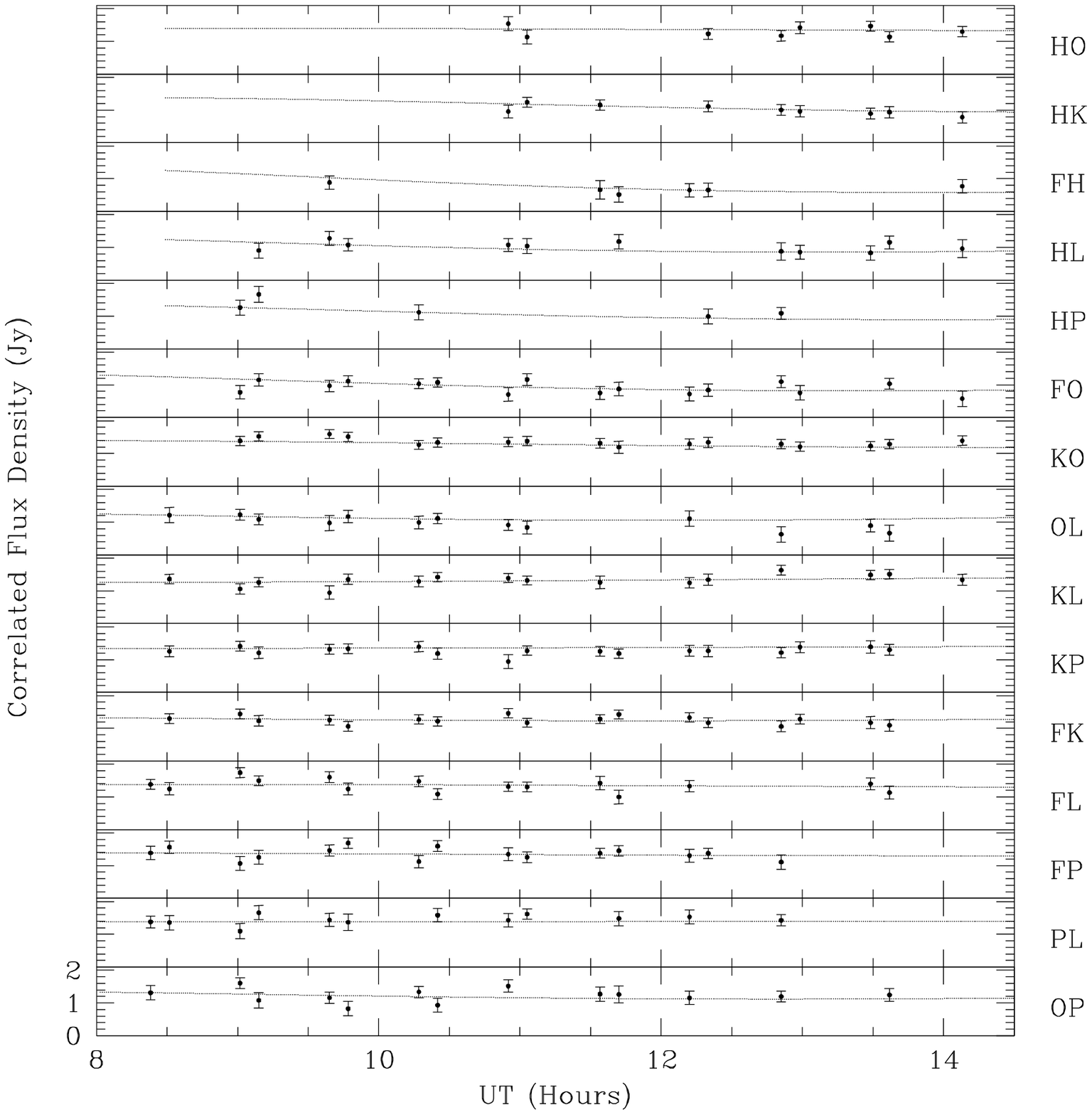,width=6.5in}
 \end{center}
\end{figure}

\begin{figure}[p]
 \begin{center}
\epsfig{figure=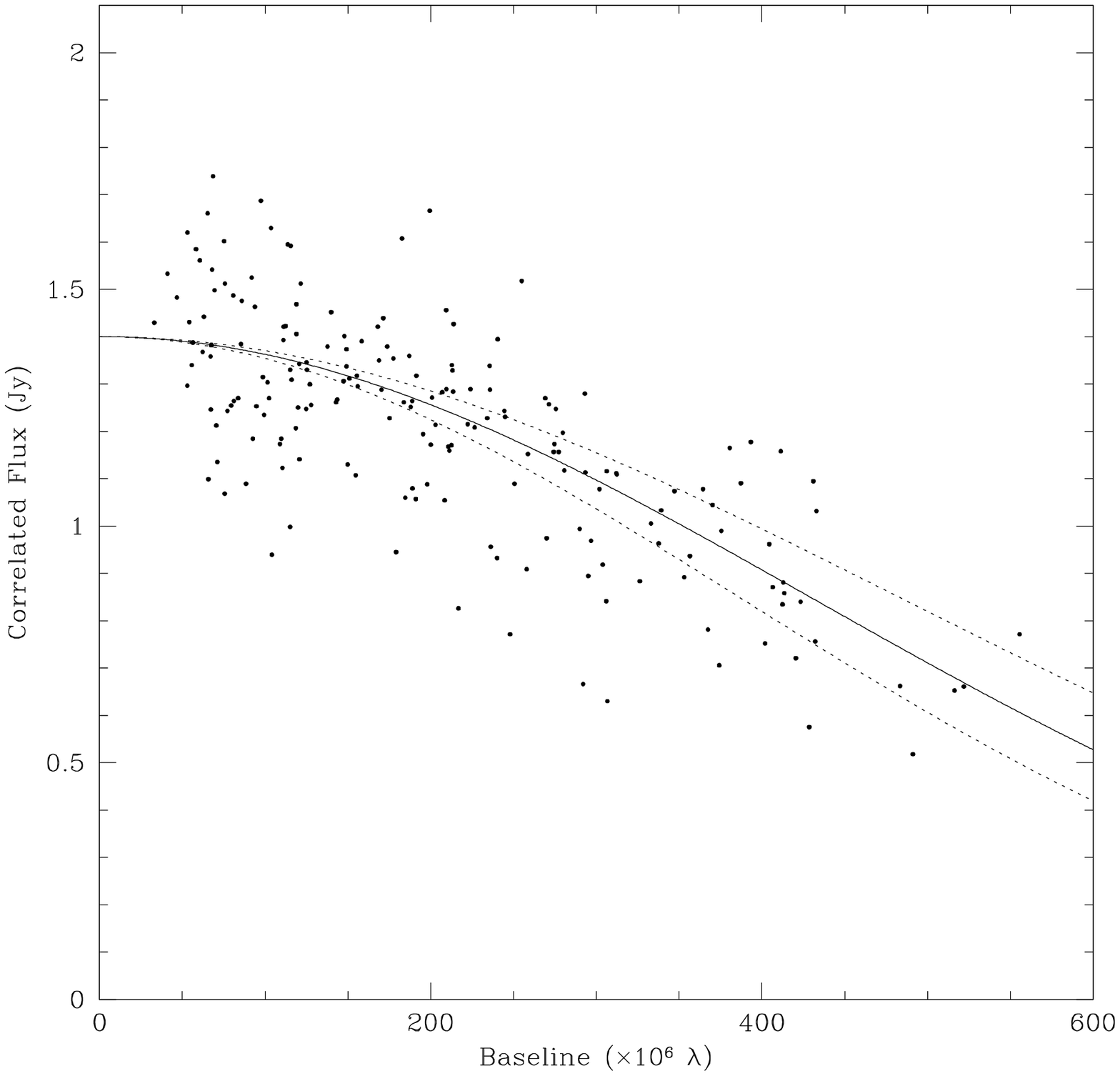,width=6.5in}
 \end{center}
\end{figure}

\begin{figure}[p]
 \begin{center}
\epsfig{figure=Doeleman.fig5.epsi,width=6.5in}
 \end{center}
\end{figure}

\begin{figure}[p]
 \begin{center}
\epsfig{figure=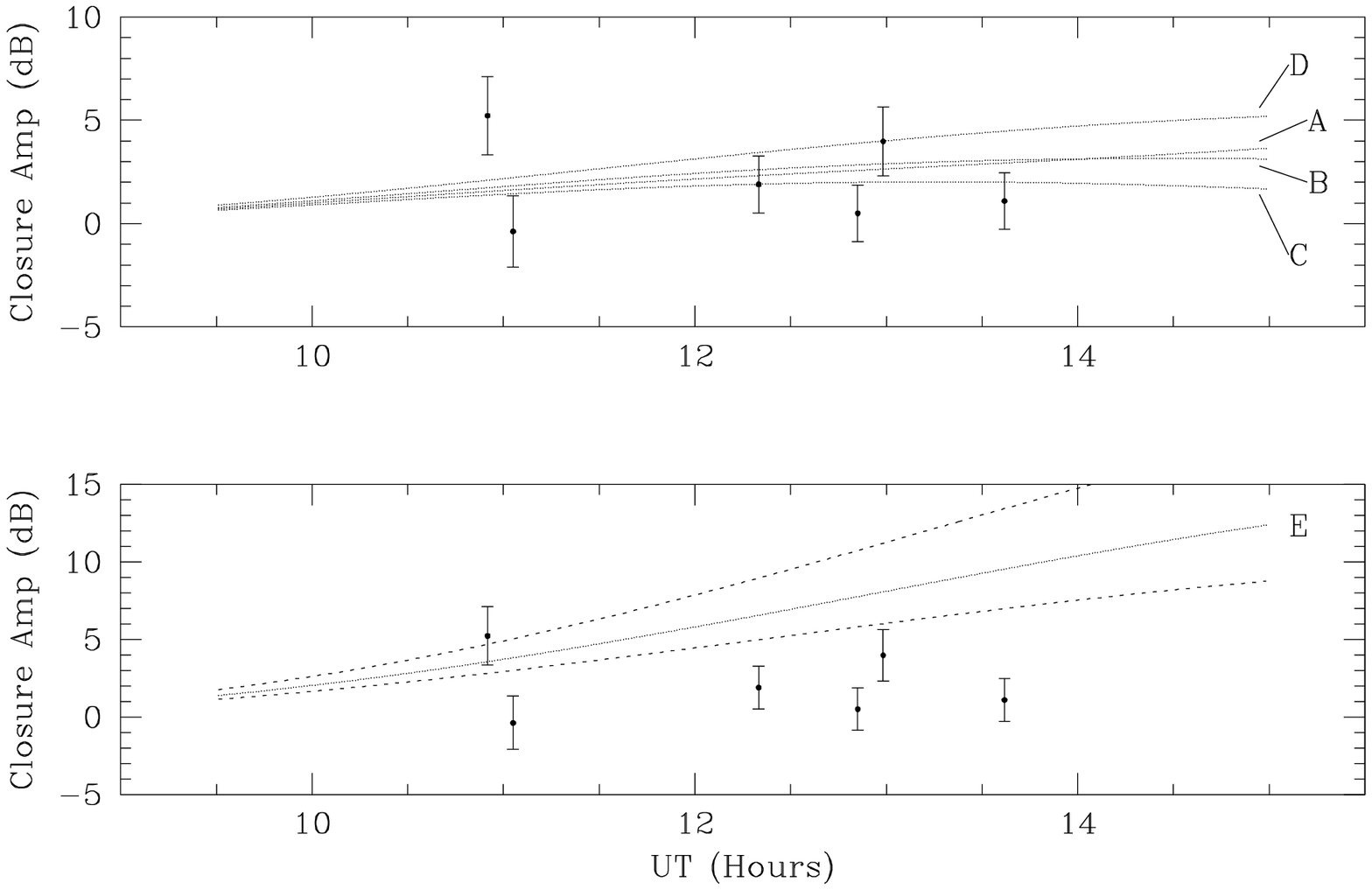,width=6.5in}
 \end{center}
\end{figure}

\begin{figure}[p]
 \begin{center}
\epsfig{figure=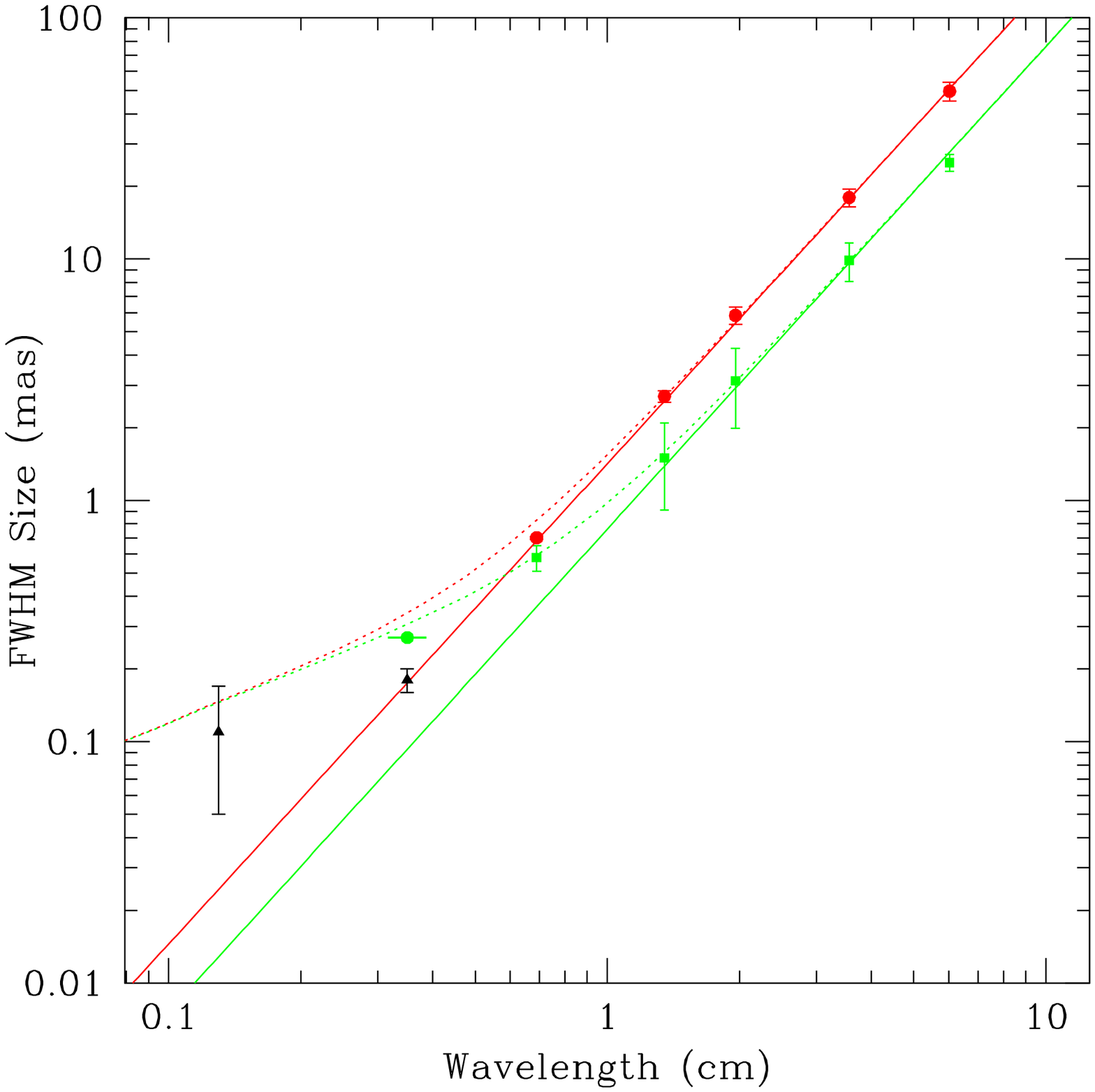,width=6.5in}
 \end{center}
\end{figure}

\end{document}